\documentclass[useAMS,usenatbib]{mn2e}
\usepackage{graphicx}

\title[The PDS vs. Markarian starburst galaxies]{The PDS vs. Markarian starburst galaxies: comparing strong and
weak \textit{IRAS} emitter at 12$\mu$m and 25$\mu$m in the nearby
universe.}
\author[R. Coziol]
  {R.~Coziol$^1$\\
  $^1$Departamento de Astronomia,
  Universidad de Guanajuato, Apartado Postal 144, 36000 Guanajuato, Gto M\'exico\\
  }
\date{Released 2002 Xxxxx XX}

\pagerange{\pageref{firstpage}--\pageref{lastpage}} \pubyear{2002}

\def\LaTeX{L\kern-.36em\raise.3ex\hbox{a}\kern-.15em
    T\kern-.1667em\lower.7ex\hbox{E}\kern-.125emX}

\begin{document}

\label{firstpage}

\maketitle

\begin{abstract}
The characteristics of the starburst galaxies from the Pico dos
Dias survey (PDS) are compared with those of the nearby UV-bright
Markarian starburst galaxies, having the same limit in redshift
($v_h < 7500$ km s$^{-1}$) and absolute $B$ magnitude ($M_B <
-18$). An important difference is found: the Markarian galaxies
are generally undetected at 12$\mu$m and 25$\mu$m in
\textit{IRAS}. This is consistent with the UV excess shown by
these galaxies and suggests that the youngest star forming regions
dominating these galaxies are relatively free of dust.

The FIR selection criteria for the PDS is shown to introduce a
strong bias towards massive (luminous) and large size late-type
spiral galaxies. This is contrary to the Markarian galaxies, which
are found to be remarkably rich in smaller size early-type
galaxies. These results suggest that only late-type spirals with a
large and massive disk are strong emitter at 12$\mu$m and 25$\mu$m
in \textit{IRAS} in the nearby universe.

The Markarian and PDS starburst galaxies are shown to share the
same environment. This rules out an explanation of the differences
observed in terms of external parameters. These differences may be
explained by assuming two different levels of evolution, the
Markarian being less evolved than the PDS galaxies. This
interpretation is fully consistent with the disk formation
hypothesis proposed in Coziol et al. (2000) to explain the special
properties of the Markarian SBNG.

\end{abstract}


\section{Introduction}
What is the origin of the nuclear star burst observed in many
nearby massive galaxies? Theoretically, interaction of galaxies
was shown to be one of the most efficient ways by which to fuel
gas in the center of a galaxy and start a burst of star formation
(Barns \& Hernquist 1996; Mihos \& Hernquist 1996). From the point
of view of observation, however, such scenario seems to fit only a
fraction of the starburst galaxies in the nearby universe. In the
Markarian sample, for example, Keel \& Van Soest (1992) have found
that only 35\% of the galaxies could be interacting with a nearby
companion. A comparable fraction of non-isolated galaxies (24\%)
was found in the MBG survey (Coziol et al. 1997). The rapid and
short duration burst predicted by many interaction models seems
also in contradiction with the relatively long period of star
formation observed in most starburst galaxies (Coziol 1996;
Goldader et al. 1997; Coziol, Doyon \& Demers 2001). These
observations suggest that other mechanisms, internal to the
galaxies, must play a role in triggering or regulating star
formation in these systems.

The effect of a bar structure was frequently proposed as a
possible burst mechanism. Theoretically, the bar was shown to be
capable of funnelling gas in the center of a galaxy and start a
nuclear burst. This hypothesis was recently tested by Consid\`ere
et al. (2000) on a sample of 16 strongly barred Markarian
starburst nucleus galaxies (SBNG). The study of the oxygen and
nitrogen gradients in these galaxies proved to be incompatible
with the expected effect of a bar. Instead, evidence was found
that the bars appeared only recently as compared to the possible
ages of the bursts. In an accompanying paper, Coziol et al.
(2000), it was shown that this phenomenon is typical of the whole
Markarian SBNG sample. Taken at face value, these observations
suggest that a large fraction of the Markarian galaxies may now be
forming a disk.

It would be highly significant to know if the characteristics of
the Markarian SBNG, as reported in Coziol et al (2000), are
typical of all massive starburst galaxies in the nearby universe.
Having previously defined a new sample of such objects, the Pico
Dos Dias (PDS) starburst galaxies (Coziol et al 1998a), a
comparison of their characteristics with those of the Markarian
SBNG seemed, therefore, like the next logical step.

The plan of the article is the following. Section~2 explains the
selection criteria used to define the two samples. The different
particularities in the Far-Infrared (FIR) and the physical
characteristics that are compared are also presented and discussed
in this section. The comparison of the two samples, supported by a
statistical analysis, takes place in Section~3. This is followed,
in Section~4, by a discussion on the possible causes of the
differences observed. The consequences of these differences for
the nature of the starburst phenomenon in nearby massive galaxies
and for galaxies at higher redshift are also discussed in this
section. A brief conclusion is presented in section~5, followed by
an appendix explaining in detail the results of the statistical
tests used for this analysis.

\section{Description of the two samples}

\subsection{Selection criteria and source of the data}

The main characteristics of the PDS starburst galaxies were
already presented and discussed in Coziol et al. (1998a). This
sample consists of 200 galaxies, selected from the \textit{IRAS}
Point Source Catalog using the following FIR criteria: 1) all the
galaxies have high or intermediate quality flux density at
12$\mu$m, 25$\mu$m, 60$\mu$m and 100$\mu$m; 2) they have an
infrared spectral index $\alpha(25, 12)$ in the range $-3.00 \le
\alpha(25, 12) \le +0.35$ and an infrared spectral index
$\alpha(60, 25)$ in the range $-2.50 \le \alpha(60, 25) \le -1.9$
(where $\alpha(\lambda1, \lambda2)=
log(S_{\lambda1}/S_{\lambda2})/log(\lambda2/\lambda1)$, and
$S_{\lambda}$ is the flux in Janskys at wavelength $\lambda$). In
Coziol et al. (1998a) it was shown that these criteria allow
distinguishing starburst from AGN galaxies with a confidence level
approaching 99\%.

The PDS starburst galaxies were found to be relatively luminous
(M$_B < -18$). The fact that the FIR criteria introduce a strong
bias against dwarf starburst galaxies (H{\sevensize \bf II}
galaxies) is not surprising, considering the low metallicity and
dust content of these galaxies (Salzer, MacAlpine \& Gordon 1988).

The physical parameters that are considered in this article were
all taken from the same source: the Lyon Meudon Extragalactic
Database (\textit{LEDA}). In \textit{LEDA}, raw data, having a
different origin, are homogenized, using specific rules, to form a
uniform catalogue(\textit{LEDA} is the basis of the \textit{RC3}
catalogue of galaxies). A complete description of the
normalization process used in \textit{LEDA} can be found in de
Vaucouleurs et al. (1991) and Paturel et al. (1997). Keeping only
galaxies for which complete information (except for kinematics) is
available in \textit{LEDA} reduces the number of PDS starburst
galaxies to 168.

The comparison sample of Markarian SBNG is composed of all the
galaxies with M$_B < -18$, having, like for the PDS, information
in \textit{LEDA}. This yields a sample of 505 galaxies. Comparing
the redshift of these galaxies with those of the PDS in Figure~1,
it is obvious that the Markarian are located at higher redshift.
To eliminate the effects of Malmquist bias, a limit on the
distance must be applied. For comparison, therefore, only galaxies
which are nearer than 7500 km s$^{-1}$ are kept in the two samples
(this corresponds to 100 Mpc, adopting H$_0=75$ km s$^{-1}$
Mpc$^{-1}$). This limit was chosen in order to exclude the
smallest number of PDS galaxies, while keeping the largest number
of Markarian galaxies. This reduce the final PDS and Markarian
samples to 154 galaxies and 325 galaxies respectively.

\begin{figure}
\includegraphics[width=200pt]{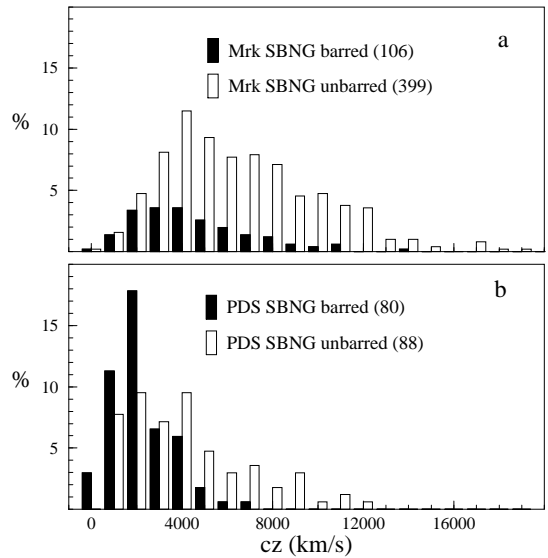}
 \caption{a) Distribution in redshift of the 505 barred and unbarred
 Markarian SBNG, with complete information in \textit{LEDA}. b) Distribution
 in redshift of the 168 PDS starburst galaxies.}
\label{fig1}
\end{figure}

\begin{figure}
\includegraphics[width=200pt]{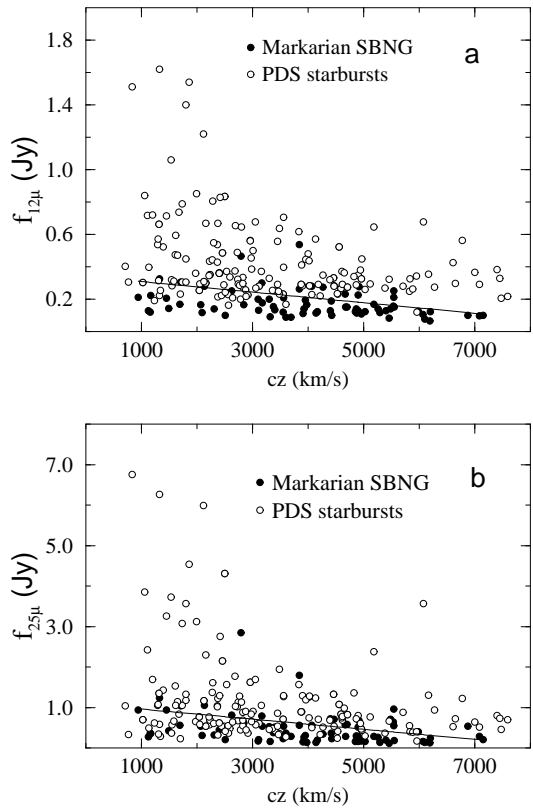}
 \caption{a) Detected flux at 12$\mu$m as a function of the redshift
 for the Markarian SBNG and PDS starburst galaxies; b) Detected flux
 at 25$\mu$m as a function of the redshift for the Markarian SBNG and
 PDS starburst galaxies. The continuous lines are linear regressions
 on the Markarian flux values. The detected fluxes for the Markarian
 SBNG are always lower than those for the PDS. Note that the slope
 of the regression goes contrary to what is expected for an
 observational bias.} \label{fig2}
\end{figure}

\begin{table*}
\begin{minipage}{420pt}
\caption{Example of the table listing the characteristics of the
PDS starburst galaxies as found in \textit{LEDA} and isolation
status as found in \textit{NED}} \label{Table1}
\begin{tabular}{lclcccccc}
\hline
\textit{IRAS}& M$_B$& Morph.&T&v$_{h}$& log(D$_{25}$)& $\mu_B$& log(V$_{max}$)& isolated\\
\hline
 & & & & (km s$^{-1}$)& (0.1 arcmin)& (mag arsec$^{-2}$)& (km s$^{-1}$)& \\
\hline
00013+2028& -20.8&  Sbc &  4.1&  2308.03&    1.54&   23.37&  2.23 &yes\\
00022-6220& -20.2&  SBbc&  3.9&  4539.00&    1.05&   22.82&       &yes\\
00073+2538& -21.5&  SBa &  1.2&  4564.64&    1.29&   22.93&  2.45 &no\\
00345-2945& -20.1&  S0/a&  0.3&  3564.14&    1.19&   23.09&  2.02 &yes\\
01053-1746& -20.8&  Irr & 10.0&  6077.20&    0.84&   21.84&  2.32 &yes\\
\hline
\end{tabular}
\medskip The complete table is available in the electronic issue of the journal.
A digital version is also available on demand to the author.
\end{minipage}
\end{table*}

\begin{table*}
\begin{minipage}{400pt}
\caption{ Example of the table listing the characteristics of the
Markarian SBNG as found in \textit{LEDA} and isolation status as
found in \textit{NED}} \label{table2}
\begin{tabular}{lclcccccl}
\hline
\textit{Mrk \#}& M$_B$& Morph.&T&v$_{h}$& log(D$_{25}$)& $\mu_B$& log(V$_{max}$)& isolated\\
\hline
 & & & & (km s$^{-1}$)& (0.1 arcmin)& (mag arsec$^{-2}$)& (km s$^{-1}$)& \\
\hline
Mrk0002  &  -20.6 & SBa  &  0.6 &  5543.45 &  0.85  & 22.06 &  2.25 & no  \\
Mrk0004  &  -20.6 & SBc  &  6.0 &  5258.60 &  1.28  & 24.14 &  2.11 & yes \\
Mrk0007  &  -19.9 & Sd   &  7.8 &  3063.23 &  0.94  & 22.01 &  2.01 & yes \\
Mrk0008  &  -20.1 & Sbc  &  4.1 &  3602.21 &  0.97  & 22.30 &  2.07 & no  \\
Mrk0011  &  -19.8 & E/S0 & -3.0 &  3909.15 &  0.96  & 22.69 &       & no  \\
\hline
\end{tabular}
\medskip The complete table is available in the electronic issue of the journal. A digital version is also available on demand to the author.
\end{minipage}
\end{table*}

\subsection{Overlap of the samples: FIR particularity of the Markarian SBNG}

Because 72\% of the Markarian galaxies were also detected in IRAS,
a large overlap between the two samples was therefore expected.
However, after comparing the galaxy names in the two samples, only
17 Markarian galaxies (5\%) were found to be common to the two
samples (this represents only 11\% of the PDS galaxies). The
reason for such low overlap is easy to understand. It is because
most of the Markarian galaxies are undetected at 12$\mu$m and/or
25$\mu$m in the \textit{IRAS} Point Source Catalogue, which is one
of the condition to be part of the PDS sample. Since this is an
important difference between the two samples, and an unrecognized
FIR characteristics for the UV-bright Markarian galaxies, a more
thorough investigation seems in order.

The reason why the Markarian SBNG are not detected at 12$\mu$m and
25$\mu$m in the Point Source Catalogue is not obvious. In part,
this is due to a lack of sensitivity. Using the \textit{IRAS}
Faint Source Catalogue instead of the Point Source Catalogue, for
instance, increases to 29\% the fraction of Markarian SBNG
detected in the four bands. However, it still leaves most of the
galaxies (71\%) as either undetected in one (33\%) or all (38\%)
of the four \textit{IRAS} bands (see Table~5 in Bicay et al.
1995).

Comparing the 12$\mu$m and 25$\mu$m fluxes of the galaxies
detected in the Faint Source Catalogue in figure~2, it can be seen
that the level of flux at these two wavebands is significantly
lower in the Markarian galaxies. The continuous lines are linear
regressions on the Markarian flux values. The slope of these
regressions go contrary to what is expected for an observational
bias, which suggests that we are still far from the observational
limits. The low detection at 12$\mu$m and 25$\mu$m looks,
therefore, as a physical trait of these galaxies: the Markarian
SBNG are generally weak FIR emitter at 12$\mu$m and 25$\mu$m.

In a sense, the above particularity may be the only one consistent
with the UV-brightness of these galaxies. The presence of a UV
excess implies that a high number of young star forming regions
present in these galaxies are relatively free of dust. If these
young and dusty free star forming regions dominate over heavily
dusty ones, then the Markarian are not expected to be particularly
bright at 12$\mu$m and 25$\mu$m, since only these young star
forming regions can heat dust sufficiently to be visible at these
two wavelengths. The lower detection rate of the Markarian SBNG at
12$\mu$m and 25$\mu$m in \textit{IRAS} marks, therefore, an
intrinsic difference between the two samples: galaxies in the
Markarian sample are either less dusty or less homogeneously
covered in dust than galaxies in the PDS sample.

This new characteristic of the Markarian SBNG is quite important.
It suggests that UV-bright and FIR bright selected samples may
differ also in other important manners. This is what this analysis
will now determine.

For comparison sake, the 17 galaxies common to the two samples
will be erased from the Markarian sample and considered as part of
the PDS sample only. Table~1 lists the characteristics of the 154
PDS starburst galaxies as found in \textit{LEDA}: column 1 gives
the \textit{IRAS} name, followed by the absolute magnitude in B,
column 2, the morphology and morphological index, $T$, column 3
and 4, the mean heliocentric redshift, column 5, the logarithm of
the apparent corrected diameter, $D_{25}$, column 6, the surface
brightness, column 7, and the logarithm of the maximum rotation
velocity, column 8. The last column indicates if the galaxy is
considered to be isolated (see section 3.2). Similar
characteristics for the Markarian SBNG are listed in Table~2.

\section{Comparison of the samples}

\subsection{Differences in mean values}

\begin{table}
\caption{Mean values and dispersion for the characteristics
compared} \label{table3}
\begin{tabular}{lcccc}
\hline
&PDS& PDS&Mrk&Mrk\\
&unbarred &barred &unbarred &barred\\
\hline
N                                       &83&71&247&61\\
$\langle$v$_{h}\rangle\ (\frac{km}{s})$ &3618&2834&4824&4129\\
std Dev.                                &1746&1406&1522&1659\\
$\langle$M$_B\rangle\ $          &-20.27&-20.54&-19.65&-20.15\\
std Dev.                         &  1.04&  0.80&  0.80&  0.82\\
$\langle\mu_B\rangle\ (\frac{mag}{arsec^{2}})$&23.01&23.11&22.99&22.94\\
std Dev.                & 0.59& 0.47& 0.69& 0.67\\
$\langle$T$\rangle\ $   &2.7&3.4&1.4&3.7\\
std Dev.                &2.9&1.5&3.7&2.1\\
$\langle$log D$_{25}\rangle\ (kpc)$ &22.47&25.33&17.18&20.64\\
std Dev.                           &10.38&11.49& 7.74& 7.53\\
$f(V_{max})$                       &74\%&87\%&35\%&74\%\\
$\langle$log V$_{max}\rangle\ (\frac{km}{s})$ &2.23&2.21&2.16&2.14\\
std Dev.                           &0.20&0.15&0.87&0.18\\
\hline
\end{tabular}
\medskip $f(V_{max})$is the fraction of galaxies with kinematics information in \textit{LEDA}.
\end{table}

The mean values for the physical parameters compared are presented
in Table~3. Since the presence or absence of a bar was shown to be
of importance in Coziol et al. (2000), the mean values were
estimated by making this distinction in each sample.

In order to evaluate if the mean values are significantly
different statistically, one-way ANOVA tests are run on all the
samples. The kind of test applied (parametric or non-parametric)
is determined by verifying if the samples have normal
distribution. Once a significant difference is observed, the usual
post-tests are performed in order to determine the origin and
nature of this difference. Complete explanations for the tests,
their interpretations and the results are presented in the
appendix.

Except for the surface brightness, $\mu_B$, all the other
parameters were found to be significantly different, at a
confidence level of 99\%. The following is a summary of what
Table~3 and the statistical post-tests (Tables~A2) reveal.

For the blue absolute magnitude, M$_B$, the unbarred Markarian
SBNG are found to be less luminous than the other galaxies. The
barred Markarian SBNG are also found to be marginally less
luminous than the PDS barred starburst galaxies.

For the morphology index, T, the unbarred Markarian SBNG are found
to be more numerous in early-type galaxies than all the other
galaxies, the difference being more pronounced compared to the
barred galaxies (Markarian and PDS). From table~3, one can note
also a large fraction of unbarred galaxies in the Markarian sample
(80\% compared to 46\% in the PDS sample).

For the velocity recession, v$_h$, it is found that the Markarian
SBNG, barred and unbarred, are located at higher redshift than the
PDS starburst galaxies. In both samples the unbarred galaxies seem
also to be marginally farther away than the unbarred ones.

For the dimension of the galaxies, D$_{25}$, the unbarred
Markarian SBNG are found to be smaller in size. A marginal
difference between the Markarian barred and PDS barred is also
observed, the former being slightly smaller than the latter. The
statistical results found for the size of the galaxies are similar
to the one observed for the luminosity.

Although the information about rotational velocity is still
incomplete (see $f($V$_{max}$) in Table~3), the statistical tests
suggest that the unbarred Markarian SBNG have smaller maximum
velocity rotation than the other galaxies. The results are
consistent with the size differences observed between the samples.

\begin{table}
\caption{Fraction of galaxies in pair or group} \label{table4}
\begin{tabular}{lccc}
\hline
                     &Barred& Unbarred&Total\\
\hline
PDS starburst        &34\%&34\%&34\%\\
Markarian SBNG       &39\%&34\%&35\%\\
\hline
\end{tabular}
\end{table}

\begin{table}
\caption{Mean values and dispersions for the samples separated by
the isolation criterion} \label{table5}
\begin{tabular}{lcccc}
\hline
&PDS& PDS&Mrk&Mrk\\
&isolated &pair/gr. &isolated &pair/gr.\\
\hline
N                                             &102   &52    &201   &107   \\
$\langle$v$_{h}\rangle\ (\frac{km}{s})$       &3256  &3258  &4728  &4606  \\
std Dev.                                      &1598  &1737  &1566  &1589  \\
$\langle$M$_B\rangle\ $                       &-20.38&-20.42&-19.71&-19.82\\
std Dev.                                      &  1.00&  0.90&  0.81&  0.85\\
$\langle\mu_B\rangle\ (\frac{mag}{arsec^{2}})$&23.03 &23.09 &23.04 &22.88 \\
std Dev.                                      & 0.54 & 0.52 & 0.66 & 0.72 \\
$\langle$T$\rangle\ $                         &2.9   &3.4   &1.9   &1.8   \\
std Dev.                                      &2.6   &1.9   &3.6   &3.6   \\
$\langle$log D$_{25}\rangle\  (kpc)$          &23.08 &25.18 &18.02 &17.57 \\
std Dev.                                      & 9.90 &12.79 & 7.88 & 7.69 \\
$f(V_{max})$                                  &76\%  &86\%  &43\%  &42\%  \\
$\langle$log V$_{max}\rangle\ (\frac{km}{s})$ &2.22  &2.22  &2.07  &2.31  \\
std Dev.                                      &0.17  &0.18  &0.20  &1.19  \\
\hline
\end{tabular}
\medskip $f(V_{max})$ is the fraction of galaxies with kinematics information in the sample.
\end{table}

\subsection{Difference of environment}

Because the environment of a galaxy may have some influence on its
evolution, it seems important to check if the two samples show any
difference on this matter. One way to do this is to compare the
fraction of galaxies in pair or group in the two samples.

Since \textit{LEDA} does not give information on the environment
of galaxies, \textit{NED} (the NASA/IPAC Extragalactic Database)
was used to search for the presence of companion galaxies in the
close environment of all the galaxies in the samples. The method
followed is identical to the one used by Campos-Aguilar \& Moles
(1991) and more recently Noeske et al. (2001). A galaxy is
considered to be non isolated when an objects is found within a
projected separation $s_p \le 0.1$ Mpc and its difference in
velocity is $\Delta v_{h} \le 500$ km s$^{-1}$. The angular search
radius was done using the mean heliocentric redshift as found in
\textit{LEDA}.

The fraction of galaxies in pair or group is presented in Table~4.
The same tendency is found for the PDS starburst galaxies and
Markarian SBNG. No significant relation is noted between the
environment and the presence of a bar.

The mean characteristics of the samples separated based on the
isolation criteria are presented in Table~5. The mean values look
more similar than when the samples are separated based on the
presence or absence of a bar.

The environment comparison is conclusive. It shows that whatever
the differences are, they cannot be explained by different
environmental effects.

\section{Discussion}

\begin{figure}
\includegraphics[width=200pt]{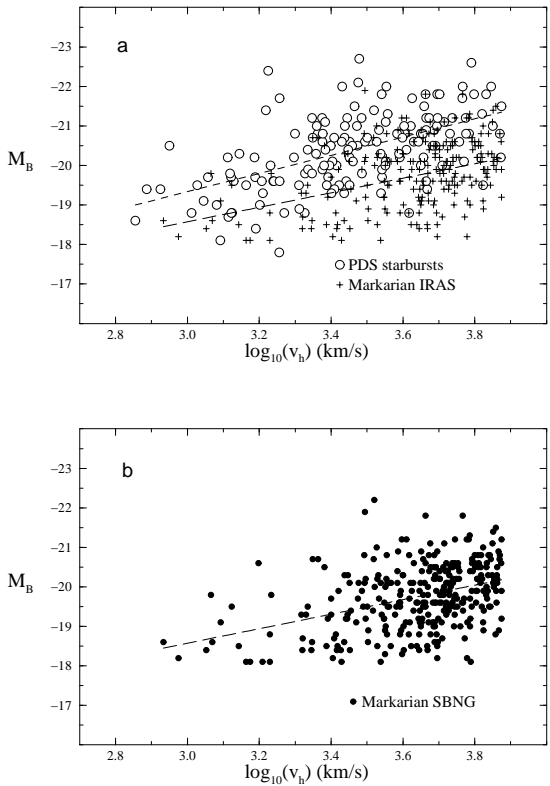}
 \caption{Absolute magnitude as a function of the redshift a)
 for the PDS starburst galaxies; b) for the Markarian SBNG.
 The dashed and long dashed curves in the two graphics are
 linear regression on the PDS and Markarian respectively.
 They show the presence of a Malmquist bias. The regression
 for the Markarian is reported in the panel a) to show the
 difference between the sample. In a) the position occupied
 by the Markarian SBNG detected in IRAS are also shown.} \label{fig3}
\end{figure}

\begin{figure}
\includegraphics[width=210pt]{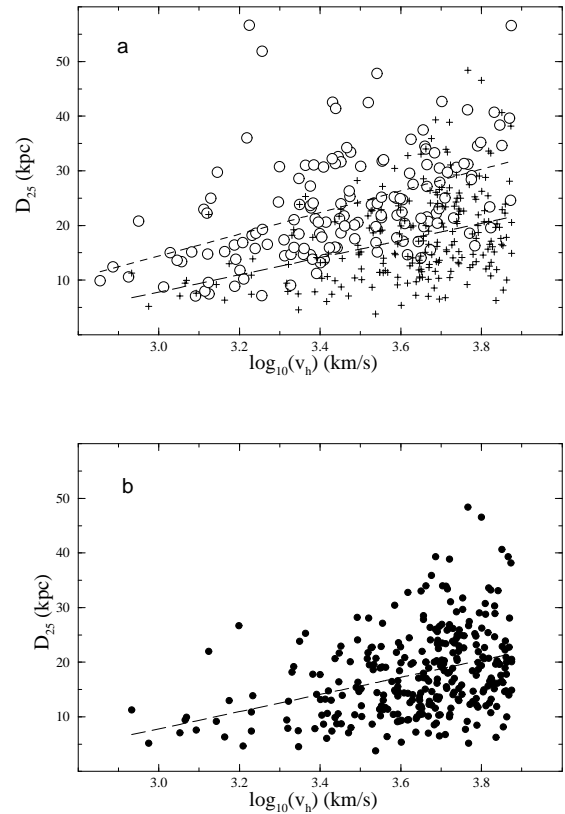}
 \caption{Diameter of the galaxies as a function of the redshift.
 a) for the PDS starburst galaxies; b) for the Markarian SBNG.
 The signification of the symbols is the same as in Figure~3.} \label{fig4}
\end{figure}

The criteria used to select the PDS seem to imply important
differences in the characteristics of the galaxies sampled. It is
important, however, to verify that these differences are not due
to spurious observational biases. Observational biases (like
Malmquist bias) are usually detected by looking for a relation
between the parameter observed and the redshift. This method was
already used in Figure~2 to show that an observational bias cannot
explain why the Markarian galaxies are not detected at 12$\mu$ and
25$\mu$ in IRAS. A bias would have cause the detected fluxes to
increase with the redshift. On the contrary, the decreasing slopes
in Figure~2 indicate the real observational limits in these bands
were not reached.

Applying the same kind of test for the absolute magnitude in
Figure~3, it can be seen that the luminosity of the galaxies is
increasing with the redshift. This is a Malmquist bias. Although
the bias seems less severe for the Markarian SBNG, the effect is
insufficient to explain the difference in luminosity between the
two samples. What Figure~3 shows also is that the PDS are more
luminous than the Markarian SBNG at all redshift. For some
reasons, therefore, the PDS sample seems to contain only the most
luminous galaxies.

\begin{figure}
\includegraphics[width=200pt]{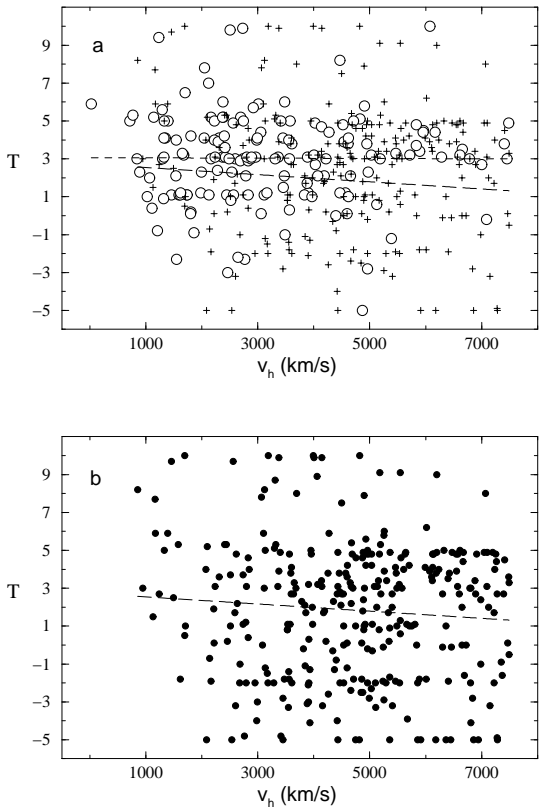}
 \caption{Morphology as a function of the redshift: a) for the
 PDS starburst galaxies; b) for the Markarian SBNG. The signification
 of the symbols is the same as in Figure~3.}
\label{fig5}
\end{figure}

The above result is consistent with the size difference observed
between the galaxies in the two samples. Assuming the luminosity
in B is correlated to the size of a galaxy, the PDS starburst
galaxies being more luminous than the Markarian SBNG are naturally
expected to have bigger size. This is verified in Figure~4. The
similarity between Figure~3 and Figure~4 (Malmquist bias included)
is obvious. It suggests that the large size of the PDS starburst
galaxies is due to a bias in luminosity. But, how is this
possible, since no condition on the B luminosity was applied?

The B luminosity bias is not trivial. In fact, the only way one
could obtain such a bias is through the FIR PDS selection
criteria. By picking galaxies that emit in the four IRAS bands,
only FIR luminous galaxies are selected. This was already noted in
Coziol et al. (1998a). Now, assuming the FIR luminosity of a
galaxy is correlated to its mass (which is consistent with the
analysis in Coziol, Doyon \& Demers 2001) and the mass is
correlated to the B luminosity, then, FIR luminous galaxies are
also expected to be luminous in B. Therefore, the B luminosity
preference observed is not an observational bias, but a result of
the FIR selection criteria used: by selecting IRAS galaxies that
emit in the four IRAS bands, we pick only FIR luminous galaxies,
which turn out to be massive (luminous in B) and large size
galaxies.

A strong indication that this interpretation is correct is the
preference in morphology shown by the PDS. The morphology of the
galaxies as a function of the redshift is examined in Figure~5. No
significant relation (bias) is observed with the redshift. It can
be seen that the PDS are mostly late-type spiral galaxies.
Late-type spiral galaxies being richer in dust than early-type
ones are obviously favoured by the FIR selection criteria.

Implicit in the definition of an observational bias, there is also
a notion of incompleteness (some objects are left aside).
Obviously, such incompleteness does not apply to the PDS sample.
The PDS is a sub-sample of the IRAS galaxy survey, which is an all
sky survey (complete up to a certain magnitude). The PDS contains,
therefore, all the nearby galaxies (non AGN) that emit in the four
IRAS bands.

The only galaxies that are missing in the present PDS sample are
those emitting in the four IRAS bands, but that were not
classified as starburst in Coziol et al. (1998a).  The properties
of the galaxies in this sample (the PDS normal galaxies) are
presented in Table~6. The same preference towards massive
(luminous) large size late-type spiral galaxies is observed. This
is independent from the fact that these galaxies are less luminous
in the FIR than the PDS starbursts ($\langle$L$_{IR}\rangle\  =
10$, compared to $\langle$L$_{IR}\rangle\  = 10.3$ for the PDS
starburst).

\begin{table}
\caption{Mean values and dispersion for the PDS normal galaxies}
\label{table6}
\begin{tabular}{lcc}
\hline
&unbarred &barred \\
\hline
N    & 74    &  83  \\
$\langle$M$_B\rangle\ $   &-20.2 &-20.4\\
$\langle\mu_B\rangle\ (\frac{mag}{arsec^{2}})$&23.1  &23.1\\
$\langle$T$\rangle\ $     &3.7    &4.2\\
$\langle$log D$_{25}\rangle\ (kpc)$       &25     &26\\
$\langle$log V$_{max}\rangle\ (\frac{km}{s})$ &2.28  &2.25\\
\hline
\end{tabular}
\end{table}

It is important to remember that the Markarian SBNG are also a
sub-sample of the IRAS galaxies. In the present sample, 72\% of
the Markarian galaxies were detected at 100$\mu$m and/or 60$\mu$m
in IRAS. These are shown with the PDS in Figure~3, 4 and 5. These
IRAS starburst galaxies are not included in the PDS sample because
they do not emit at 12$\mu$m and 25$\mu$m. If we relax this
criterion the two samples merge together and we loose all
distinction between the PDS and Markarian starburst galaxies. We
then have a sample of IRAS starburst galaxies covering all ranges
in luminosity, size and morphology.

From the above analysis, it is concluded that there is a strong
correlation between the physical characteristics of the galaxies
sampled and the fact that they are weak or strong emitter at
12$\mu$m and 25$\mu$m in IRAS. It seems therefore secure to state
that the only galaxies that emit in these bands in the nearby
universe are massive (luminous) large size late-type spiral
galaxies. Equivalently, we can also state that the reason the
Markarian SBNG are not detected in these bands is because they are
too small and have too early morphology.

Note that the high number of early-type galaxies in the Markarian
sample may already explain the large difference in barred galaxies
observed in the different samples (see Table~3). In the Markarian
sample, 111 of the unbarred galaxies are early-type ($T\leq0$).
Since bars are less frequent (or more difficult to detect) in
these galaxies, their high number in this sample leads naturally
to a lower frequency of barred galaxies. Eliminating the 111
early-type unbarred galaxies, for example, increases the fraction
of barred galaxies to 31\%, which is still low, but more
comparable to the 46\% barred galaxies observed in the PDS sample.

One important question is why are the Markarian SBNG so numerous
in early-type galaxies? One possibility is to assume a special
burst mechanisms. Merger of galaxies is usually recognized as a
mechanism capable of producing early-type galaxies. This mechanism
may also be fatal for a pre-existing spiral disk and a bar. If
merger is the main burst mechanism of the Markarian SBNG, a high
number of early type galaxies in this sample is thus
understandable.

In the case of the PDS galaxies, their different properties may be
the result of gas accretion. In Coziol et al. (1998b), using a
chemical evolution model, it was shown that a starburst galaxy
that is accreting gas forms a late-type spiral with a metallicity
that is high as compared to a merger case. Having a high
metallicity, these galaxies are expected to be rich in dust and to
possibly emit in the four IRAS bands.

One difficulty with the above explanation is that nothing
indicates what could trigger one mechanism instead of the other.
The fact that the galaxies in the two samples were shown to share
the same environment excludes an explanation in terms of external
causes.

Another possibility is to assume that we are observing a unique
population of starburst galaxies where galaxies are observed at
different level of evolution. As previously shown, this would be
equivalent to relaxing the FIR criteria used to select the PDS.
The problem now is to identify the process that corresponds to the
evolutionary sequence observed.

To explain the characteristics of the Markarian SBNG, it was
proposed in Coziol et al. (2000) that they are presently forming a
disk. Assuming this process is not instantaneous, we should then
expect to find, in a single population, examples of galaxies at
different stage of formation of their disk. According to this
interpretation, the Markarian SBNG are simply at a less advanced
stage than the PDS starbursts.

One could easily test the above hypothesis. It would be sufficient
to determine if the PDS have different metallicity and metallicity
gradients (O/H and N/O) than the Markarian SBNG (Consid\`ere et
al. 2000). Assuming the PDS are more evolved, a strong influence
of the bar would also be expected in these galaxies.

In Coziol et al. (2000), the proposition that the Markarian
galaxies are in the process of forming a disk was said to be
consistent with observations of galaxies in the HDF, suggesting
that large size spiral galaxies become less frequent at higher
redshift (van den Bergh et al. 2000). However, assuming the same
process happens for galaxies in the HDF, we would then expect to
also see a difference in size between the early-type and late-type
galaxies in this sample. The fact that this difference was
recently observed (Cohen et al. 2003) may be one new argument in
support of the disk formation model.

\section{conclusion}

The goal of this article was to verify if other nearby starburst
galaxies share the characteristics noted for the Markarian SBNG in
Coziol et al. (2000). Using the PDS starburst galaxies, which were
selected based on FIR criteria, it was found that the type of
galaxy detected differs quite significantly. These differences
were then shown to be strongly correlated to the fact that the PDS
are strong emitter at 12$\mu$m and 25$\mu$m in IRAS. In
particular, it suggests that only massive and large size late-type
spiral galaxies in the nearby universe have sufficiently hot dust
to be observed in these two bands.

The differences observed between the two samples are not in
contradiction with the interpretation proposed in Coziol et al.
(2000), which is that a large number of Markarian SBNG are in the
process of forming a disk. If such process extends over a
significant period of time, it is naturally expected to find, in a
single population, examples of galaxies at different stages of the
formation of their disk. The characteristics expected for galaxies
at a more advanced stage are similar to those observed for the
PDS. This interpretation is consistent with the observation that
the PDS starburst and Markarian SBNG share the same environment.

Similarities with what is observed in the HDF (van den Bergh et
al. 2000; Cohen et al. 2003) suggests that we may see the same
phenomenon at high and low redshift. This supports the view that
nearby starburst in massive galaxies are examples of galaxy in
formation.

\thanks
The author thanks the referee for important comments. This
research was supported by CONACyT grant EX-000479. For this
analysis, the Lyon-Meudon Extragalactic Database (\textit{LEDA}),
operated by the Lyon and Paris-Meudon Observatories (France) and
the NASA/IPAC Extragalactic Database (\textit{NED}) which is
operated by the Jet Propulsion Laboratory, California Institute of
Technology, under contract with the National Aeronautics and Space
Administration were used.

\appendix

\section{One-way analysis of variance: description and results}

The one-way analysis of variance (ANOVA) test compares three or
more unmatched groups of data, based on the assumption that the
two populations are Gaussian. The P value answers this question:
if the populations really have the same mean, what is the chance
that random sampling would result in means as far apart as
observed?

The post tests (like Tukey's and Dun's post tests) are
modifications of the t test. They account for multiple
comparisons, as well as for the fact that the comparisons are
interrelated. The unpaired t test computes the t ratio as the
difference between two group means divided by the standard error
of the difference (computed from the standard errors of the two
group means, and the two sample sizes). The P value is then
derived from t. The post tests work in a similar way. Instead of
dividing by the standard error of the difference, they divide by a
value computed from the residual mean square. Each test uses a
different method to derive a P value from this ratio.

For the difference between each pair of means, the tests report
the P value as $>0.05$, $<0.05$, $<0.01$ or $<0.001$. If the null
hypothesis is true (all the values are sampled from populations
with the same mean), then there is only a 5\% chance that any one
or more comparisons will have a P value less than 0.05.

For comparison the data for the PDS and Markarian starbursts were
separated in 4 groups: group A, the PDS unbarred galaxies, group
B, the PDS barred, group C, the Markarian unbarred and group D,
the Markarian barred. Before performing the analysis,
Kolmogorov-Smirnov tests were used to verify the normality
distribution of the data. The only group of data that did not pass
this test is the maximum rotation velocity for the Markarian
unbarred galaxies (group C). A non-parametric test
(Kruskall-Wallis) and Dunn's post test were used for this
parameter.

Results of the one-way ANOVA and non-parametric tests are reported
in Table~A1. The results for the post tests, when the means were
shown to varied significantly, are reported in
Table~A2.\footnote{For this analysis, the program Prism, version
3.00, for Windows was used. This program was made by GraphPad
software, San Diego California USA (www.graphpad.com).}

\begin{table}
\caption{Result for the one-way ANOVA and non parametric tests}
\label{TableA1}
\begin{tabular}{lcc}
\hline
Parameters& P&Means signif. different\\
\hline
M$_B$& $<0.001$&yes\\
$\mu_B$& 0.4637&no\\
T& $<0.001$&yes\\
v$_{h}$& $<0.001$&yes\\
D$_{25}$& $<0.001$&yes\\
V$_{max}$&$<0.001$&yes\\
\hline
\end{tabular}
\end{table}

\begin{table}
\caption{P values for the post tests} \label{TableA2}
\begin{tabular}{lccccc}
\hline
groups & M$_B$  & T      & v$_{h}$&D$_{25}$& V$_{max}$\\
\hline
A vs B &$>0.05$ &$>0.05$ &$<0.05$ &$>0.05$ &$>0.05$\\
A vs C &$<0.001$&$<0.01$ &$<0.001$&$<0.001$&$<0.001$\\
A vs D &$>0.05$ &$>0.05$ &$>0.05$ &$>0.05$ &$>0.05$\\
B vs C &$<0.001$&$<0.001$&$<0.001$&$<0.001$&$<0.001$\\
B vs D &$<0.05$ &$>0.05$ &$<0.001$&$<0.05$ &$>0.05$\\
C vs D &$<0.001$&$<0.001$&$<0.05$ &$<0.05$ &$>0.05$\\
\hline
\end{tabular}
\end{table}

\label{lastpage}

\end{document}